\documentclass[12pt]{article}

\usepackage{amsmath}
\usepackage{amssymb}
\usepackage{authblk}
\usepackage{color}
\usepackage{graphicx}
\usepackage{caption}
\usepackage{subcaption}

\def\L*{{\cal L}_*}
\def\L{\mathcal{L}}

\def\lsim{\mathrel{\rlap{\lower3pt\hbox{\hskip0pt$\sim$}}
     \raise1pt\hbox{$<$}}}         
\def\gsim{\mathrel{\rlap{\lower4pt\hbox{\hskip1pt$\sim$}}
     \raise1pt\hbox{$>$}}}         

\def\opsi{\hat{\psi}(\theta)}
\def\opsid{\hat{\psi}^\dagger(\theta)}
\def\Ngs{\ket{0_N}}

\newcommand{\bra}[1]{\langle{#1}|}
\newcommand{\ket}[1]{|{#1}\rangle}
\newcommand{\braket}[2]{\langle{#1}|{#2}\rangle}

\DeclareMathOperator*{\tr}{Tr}

\title{Black Holes and Quantumness on Macroscopic Scales}
\author{D. Flassig}
\author{A. Pritzel}
\author{N. Wintergerst}
\affil{Arnold Sommerfeld Center for Theoretical Physics, Fakult\"at f\"ur
Physik, \authorcr
Ludwig-Maximilians-Universit\"at M\"unchen, Theresienstr.~37, 80333 M\"unchen,
Germany} 
\date{Dec. 13th 2012}

\begin{document}
\maketitle

\begin{abstract}
It has recently been suggested that black holes may be described as
condensates of weakly interacting gravitons at a critical point, exhibiting strong quantum effects.
In this paper, we study a model system of attractive bosons in one spatial dimension which is known to undergo a quantum phase transition.
We demonstrate explicitly that indeed quantum effects are important at the critical point, even if the number of particles is macroscopic.
Most prominently, we evaluate the entropy of entanglement between different momentum modes and observe it to become maximal at the critical point.
Furthermore, we explicitly see that the leading entanglement is between long wavelength modes and is hence a feature independent of ultraviolet physics.
If applicable to black holes, our findings substantiate the conjectured breakdown of semiclassical physics even for large black holes. This can resolve long standing mysteries, such as the information paradox and the no-hair theorem. 
\end{abstract}

\section{Introduction} 

In a recent series of papers \cite{nport, lglim, hair, critpoint}, Dvali and Gomez have proposed a new 
conceptual framework for the understanding of black hole physics. The first of
two key claims is that the black hole is a bound state or condensate of many weakly interacting (i.e. long-wavelength) 
gravitons. Secondly, it was suggested that this condensate is at a quantum critical point and therefore 
exhibits properties that are not apparent in the traditional description in terms of (semi-)classical general
relativity. Most importantly, the underlying quantum physics could be able to
resolve the mysteries of the information paradox. Hawking evaporation is described as the depletion and 
evaporation of the condensate and its purification is thus a natural result. 
To the same extent, black holes could carry quantum hair \cite{hair}. 
These effects are not visible in the semiclassical approximation, since this
limit corresponds to an infinite number of black hole constituents.

In this letter, we will review the Dvali-Gomez
proposal and will elaborate in more detail why one can expect Bose
condensates at a critical point to display qualitatively new phenomena.
In particular, we will discuss how quantum physics can be relevant on macroscopic scales in such
systems. To this end, we are going to investigate in detail the
quantum phase transition of the attractive Bose gas in $1+1$ dimensions.

The idea that black holes may be described by the quantum physics of $N$ weakly
interacting gravitons was first put forward in \cite{nport}. There it was
observed that in pure Einstein gravity a black hole of mass $M$ and hence size
$R_s = G_N M$ primarily consists of gravitons of wavelength $\lambda \sim R_s$. 
As each of these long wavelength particles contributes energy $E_1 = \hbar /
\lambda$, one obtains
\begin{equation}
\label{Ndef} N \sim \frac{M}{E_1} = \frac{M^2}{M_p^2} .
\end{equation}

The idea of a black hole as a Bose condensate of gravitons can also be motivated in a 
bottom-up approach. As gravitons are self coupled, they can potentially form a self 
sustained bound state. The properties of such a bound state can be estimated via
the virial theorem, 
\begin{equation}
	\langle E_\text{kin} \rangle \sim \langle V \rangle \, . 
\end{equation}
The kinetic energy $E_\text{kin}$ of $N$ gravitons of wavelength $\lambda$ is given by 
\begin{equation}
\langle E_\text{kin} \rangle = N \frac{\hbar}{\lambda} \, , 
\end{equation}
while a naive estimate of the potential energy of the configuration of size $R$
is
\begin{equation}
	\langle V \rangle \sim N^2 \frac{G_N \hbar^2}{\lambda^2 R} \,. 
\end{equation} 
Assuming the size to be of the order of the wavelength, $R \sim \lambda$, one
obtains
\begin{equation}
	\lambda \sim \sqrt{N} L_p \, . 
\end{equation}
It is easily verified that this relation is nothing but Eq. (\ref{Ndef}).
An order of magnitude estimate of graviton-emission gives a result
consistent with the rate of emission of Hawking radiation. Consequently, a self
sustained bound state of gravitons, if it exists, will likely behave like a
black hole.

One should observe that the typical interaction strength between two gravitons 
is $\alpha \sim 1/(\lambda M_{Pl})^2$. However, all mutual 
interactions add up and their total effect should be quantified by $N \alpha$. This implies
that the self bound graviton condensate is at $\alpha N\sim 1$,
where interactions start to dominate over the kinetic term. This condition
characterizes the critical point of a zero temperature phase transition or
quantum phase transition (QPT) in a simple bosonic model system
\cite{critpoint} and is considered to ensure that quantum effects are important
even for macroscopic black holes.

To substantiate this idea, it is our interest to gain more qualitative
insight into bosonic systems at a critical point by a detailed study of this
$1+1$-dimensional nonrelativistic attractive Bose gas on a ring. The transition in this system was
discovered and first studied in \cite{kanamoto_etal}. We will substantiate the existence of this critical point
by studying appropriate characteristics.

We will then focus on the quantum behavior. As a measure of quantumness, we
calculate the entanglement of different momentum modes applying analytical as
well as numerical techniques. We observe that it becomes maximal at the critical
point and for low momentum modes. We interpret this as further evidence that the
black hole condensate picture can be successful independent of the ultraviolet
physics that completes Einstein theory.

The remainder of the paper will be organized as follows. In section 2 we
will introduce in detail the $1+1$-dimensional attractive Bose gas, remind the
reader of mean g  and introduce the basis of our numerical studies. 
Further evidence for the existence of a quantum critical point is provided in
section 3. We will then introduce the fluctuation entanglement as a
relevant measure of quantumness and present our results in 4. Finally, in
the conclusions, we discuss the qualitative consequences of our findings with 
regards to the physics of macroscopic black holes.

\section{The $1+1$-dimensional Bose Gas}

Throughout this paper, we consider a Bose gas on a $1D$-circle of radius $R$ with attractive interactions. The Hamiltonian is given by
\begin{equation}
	\label{hamiltonian}
	\hat{H} = \frac{1}{R} \int_0^{2\pi} d\theta \left[-\frac{\hbar^2}{2m}\opsid \partial_\theta^2 \opsi - \frac{\hbar^2}{2m} \frac{\pi \alpha R}{2}\opsid\opsid\opsi\opsi \right]\,,
\end{equation} 
where $\alpha$ is a dimensionless, positive coupling constant.
This Hamiltonian can be cast into a more convenient form by decomposing $\opsi$ in terms of annihilation operators:
\begin{equation}
	\label{psidec} 
	\opsi = \frac{1}{\sqrt{2\pi R}}\sum_{k=-\infty}^{\infty} \hat{a}_k e^{i k \theta} \,,
\end{equation}
which leads to 
\begin{equation}
\label{momhamilt}
\hat{H} = \sum_{k=-\infty}^\infty k^2 \hat{a}_k^\dagger \hat{a}_k - \frac{\alpha}{4}\sum_{k,l,m = -\infty}^\infty \hat{a}_k^\dagger \hat{a}_l^\dagger \hat{a}_{m+k} \hat{a}_{l-m}
\end{equation}
Note that in order to improve readability we have now switched to units $R = \hbar = 2m = 1$. 
The total number operator is 
\begin{equation}
\hat{N} = \int_0^{2\pi} d\theta \opsid\opsi = \sum_{k=-\infty}^{\infty} \hat{a}_k^\dagger \hat{a}_k \,.
\end{equation}
It was first shown in \cite{kanamoto_etal} that an increase of the effective coupling $\alpha N$ on the ring leads to a transition from a homogenous ground state to a solitonic phase, where the critical point is reached for $\alpha N = 1$. 

\subsection{Mean Field Analysis}

A mean field approach to the hamiltonian (\ref{hamiltonian}) leads to the Gross-Pitaevskii energy functional
\begin{equation}
\label{gpefunc}
  E[\Psi_{GP}] = \int_0^{2\pi} d\theta \left[\left|\partial_\theta \Psi(\theta)\right|^2 - \frac{\alpha}{2}|\Psi(\theta)|^4 \right]
\end{equation}
The ground state wavefunction $\Psi_0$ is obtained through minimization of the energy functional subject to the constraint $\int d\theta |\Psi(\theta)|^2 = N$. This leads to the time independent Gross-Pitaevskii equation 
\begin{equation}
  \left[\partial_\theta^2 + \pi \alpha |\Psi_0(\theta)|^2\right]\Psi_0(\theta) = \mu \Psi_0(\theta) \,,
\end{equation}
where $\mu = dE/dN$ is the chemical potential. Solutions to this equation are given by (see e.g. \cite{carr})\footnote{Here, ${\rm dn}(u|m)$ is a  Jacobi elliptic function and $K(m)$ and $E(m)$ are the complete elliptic integrals of the first and second kind, respectively.}
\begin{equation}
\label{soliton}
  \Psi_0(\theta) = 
  \begin{cases} 
  \sqrt{\frac{N}{2\pi}} & \mu = \frac{\alpha N}{2} \\
  \sqrt{\frac{N K(m)}{2\pi E(m)}} {\rm dn}\left(\frac{E(m)}{\pi}(\theta-\theta_0)|m\right) & \mu = \frac{(2-m)K^2(m)}{\pi^2}
  \end{cases} \,.
\end{equation}
Here, $\theta_0$ denotes the center of the soliton and $m$ is determined by the
equation
\begin{equation}
	K(m)E(m)=\left(\frac{\pi}{2}\right)^2 \alpha N \,.
\end{equation}

For small $\alpha N<1$, (\ref{gpefunc}) is minimized by the homogenous wavefunction. On the other hand, for $\alpha N>1$ the solitonic solution has a lower energy. At $\alpha N = 1$, both configurations are degenerate in energy - a clear indication for a quantum phase transition.

On a side note, one may wonder whether the one-soliton solution is stable for arbitrary $\alpha N > 1$ or if multi-soliton solutions may eventually be energetically favored. This can be checked in a simple argument.
A soliton of size $R_s$ has a total energy
\begin{equation}
E \sim \frac{N}{R_s^2} - \alpha\frac{N^2}{R_s} \,.
\end{equation}
Minimization with respect to $R$ yields $R_s = \frac{2}{\alpha N}$ and
$E_1 = -\frac{1}{4}\alpha^2 N^3$. A split into two stable solitons of boson number $r N$ and $(1-r)N$ yields a total energy $E_2 = -\frac{1}{4}\alpha^2 N^3[1 - 3r(1-r)]$. This is bigger than $E_1$ for any $r < 1$. 
This can be straightforwardly generalized two multi-soliton solutions; therefore, the single soliton is stable.

Finally, let us note that the apparent spontaneous breaking of translation symmetry in the solitonic phase is in no contradiction to known theorems about the absence of finite volume symmetry breaking. The Gross-Pitaevskii ground state only becomes exact in the $N \to \infty$ limit. In this limit, translated Gross-Pitaevskii states are orthogonal and do not mix under time evolution. Technically, symmetry breaking is made possible because expecation values of composite operators made out of the fields diverge in the large $N$ limit. We comment on this in more detail in the Appendix.
 
This again emphasizes how the classical limit really emerges as a large $N$
limit from quantum mechanics. Exactly how this argument breaks down at the
critical point and what the implications of this breakdown are will be the focus
of the remainder of this manuscript.

\subsection{Bogoliubov Approximation}
The Gross-Pitaevskii equation is the zeroth-order equation in an expansion of the field operator into its mean value and quantum (and, in more general setups, thermal) fluctuations around it:
\begin{equation}
  \opsi =  \langle\opsi\rangle + \delta\hat\psi(\theta) \,.
\end{equation}
The spectrum of these small excitations around the mean field can then be found in the Bogoliubov approximation. Generally, this corresponds to approximating the fluctuation Hamiltonian by its quadratic term and subsequent diagonalization through canonical transformations of the field.  

For $\alpha N < 1$, i.e. on the homogeneous background, it is convenient to
stick to the momentum decomposition (\ref{psidec}) and replace $\hat a_0=\hat
a_0^\dagger=\sqrt{N_0} \sim \sqrt{N}$. In words, one assumes that the zero mode
is macroscopically occupied and all commutators $[\hat a_0,\hat a_0^\dagger]$ in
the Hamiltonian are suppressed by relative powers of $1/N$; the quantum
fluctuations of the zero mode may therefore be neglected. This, in combination with taking into account the constraint
\begin{equation}
\hat N = N_0 + \sum_{k \neq 0} \hat a_k^\dagger \hat a_k
\end{equation} 
leads to the Hamiltonian
\begin{equation}
	{\mathcal H}=\sum_{k\neq 0} \left(k^2-\alpha N/2\right)a_k^\dagger a_k-\frac{1}{4}\alpha N \sum_{k \neq 0}\left(a_k^\dagger a_{-k}^\dagger +a_k a_{-k}\right)+{\cal O}(1/N) \,.
\end{equation}
All interaction terms are suppressed by $1/N$ and go to zero in the double scaling limit $N \to \infty$, $\alpha \to 0$ with $\alpha N$ finite. The Hamiltonian can be diagonalized 
\begin{equation}
	{\cal H}=\sum_{k \neq 0}\epsilon_k b^\dagger_k b_k,\:
	\epsilon_k=\sqrt{k^2(k^2-\alpha N)}
\end{equation}
with a Bogoliubov transformation
\begin{equation}
	\label{bog-hom}
	a_k=u_k b_k +v^\star_k b_{-k}^\dagger,
\end{equation}
where the Bogoliubov coefficients are
\begin{align}
	u^2_{k} &=\frac{1}{2}\left[1+\frac{k^2-\frac{\alpha N}{2}}{\epsilon_k}\right],&
	v^2_{k} &=\frac{1}{2}\left[-1+\frac{k^2-\frac{\alpha N}{2}}{\epsilon_k}\right].
\end{align}
The Boguliubov approximation breaks down whenever an $\epsilon_k$ becomes too small. In that case the initial assumption that only the zero mode is macroscopically occupied is no longer justified. Obviously, it is $\epsilon_1$ that first goes to zero, namely when $\alpha N \to 1$. Right at the phase transition, the Boguliubov approximation is never valid. It is worth noting however, that for any finite distance $\delta$ from the critical point, there exists a minimal $N$ for which the approximation is valid. In other words, for any finite $\delta$, the Boguliubov approximation becomes exact in the limit $N \to \infty$. This is due to the fact that both the interaction terms as well as $v_k^2/N$ vanish in this limit for any finite $\delta$. For $\delta = 0$, however, this is never true.

In the $\alpha N > 1$ case, the classical background is not homogenous any more, but is given by the bright soliton solution (\ref{soliton}).
In this case, the background induces an additional nontrivial mixing between momentum
eigenmodes of different $|k|$. A decomposition into momentum eigenmodes requires an (unknown) analytic expression for the Fourier components of the soliton and is thus no longer convenient. On the other hand, an analytic Bogoliubov treatment is still
possible by directly decomposing $\delta\hat\psi$ into normal modes: 
\begin{equation}
\label{bogdgtrafo}
	\delta\hat\psi(\theta) = \sum_i \left(u_i(\theta) \hat b_i^\dagger +
	v_i^\star(\theta) \hat b_i \right) \,.
\end{equation}
If the mode functions obey the Boguliubov-de Gennes equations
\begin{eqnarray}
\partial_\theta^2 u_j + \alpha \Psi_0^2(2 u_j + v_j) + \mu u_j &=& E_j u_j \\
\partial_\theta^2 v_j + \alpha \Psi_0^2(2 v_j + u_j) + \mu u_j &=& -E_j u_j
\end{eqnarray}
and are normalized such that they form a complete set and the transformation (\ref{bogdgtrafo}) is canonical, the Hamiltonian is diagonalized.
The first excited Bogoliubov modes have the form
\begin{align}
	u_1(\theta)&=N_1 {\rm sn}^2\left(\left.\frac{K(m)}{\pi}(\theta-\theta_0)\right| m\right)\\
	v_1(\theta)&=-N_1 {\rm cn}^2\left(\left.\frac{K(m)}{\pi}(\theta-\theta_0)\right| m\right).
\end{align}
 The coefficient $N_1$ is defined by
\begin{equation}
	N_1^2=\frac{mK(m)}{2\pi\left[(2-m)K(m)-2E(m)\right]} \,.
\end{equation} 

\subsection{Numerical Diagonalization}
While the Bogoliubov treatment provides an approximative description of the Bose gas deep in the respective phases, it fails, as we have reasoned above, around the critical point.  

A complementary method to explore the quantum properties of the system is
numerical diagonalization of the Hamiltonian. Of course, numerical techniques
are only applicable for sufficiently small Hilbert spaces.
The Hamiltonian (\ref{hamiltonian}) is number conserving. This allows for exact diagonalization of (\ref{momhamilt}) by considering a subspace of fixed $N$. However, to make any numerical procedure feasible, we need to limit the allowed momenta. In the spirit of \cite{kanamoto_etal}, we truncate the basis of free states in which we perform the diagonalization to $|l| = 0,1$. This gives a very good approximation to the low energy spectrum of the theory well beyond the phase transition. Analytically, this can be seen by analyzing the spectrum of the soliton solution (\ref{soliton}). Only for $\alpha N \gsim 1.5$, higher $l$ modes start giving relevant contributions. We have further verified this numerically by allowing for $|l| = 2,3$; the low energy modes are only marginally affected up until $\alpha N \sim 2$.
Our code allowed us to consider particle numbers $N \lsim 10000$. In order to illustrate scaling properties, all analyses are performed for various particle numbers. 
  
\begin{figure}[t]
\centering
\includegraphics[width=.75\linewidth]{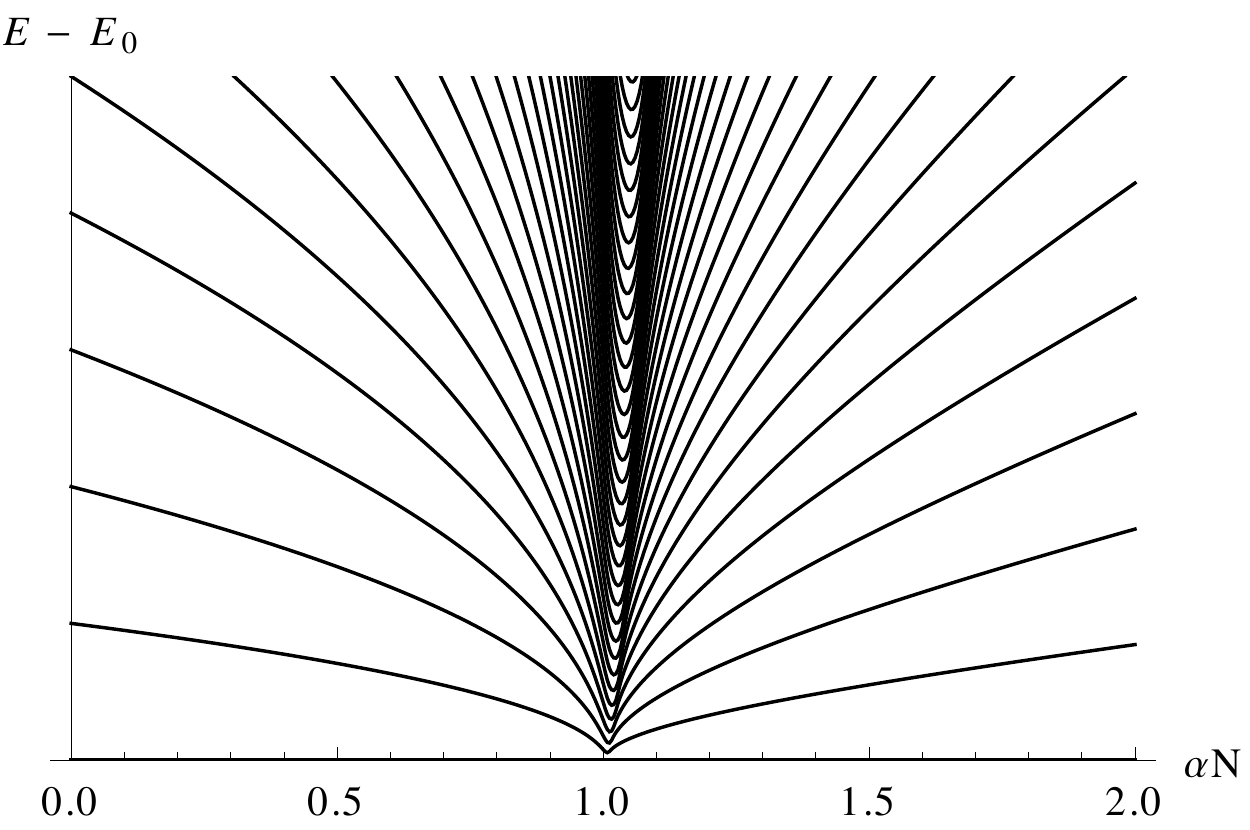}
\caption{Energy spectrum for $N = 5000$ as a function of the effective coupling $\alpha N$ }
\label{fig:spectrum}
\vspace{0.5cm}
\end{figure}

 Since the normalized coupling $\alpha N$ is the relevant quantity for a phase transition, one can analyze all interesting properties for a fixed $N$ by varying $\alpha$. The corresponding spectrum of excitations above the ground state as a function of $\alpha N$ is shown in Fig.\ref{fig:spectrum} for $N = 5000$ and $-1 \leq k \leq 1$.
One observes a decrease in the energy gap between the low lying excitations due to the attractive interactions as $\alpha N$ is increased. At the quantum critical point, the spacing between levels reaches its minimum. Its magnitude depends on the particle number $N$; 
the energy of the lowest lying excitation decreases with $N$.
By further increasing the coupling $\alpha$ one reaches the solitonic phase. The spectrum corresponds to that of translations and deformations of a soliton.

Obviously, (\ref{hamiltonian}) is invariant under translations; since we are considering a finite length ring, the ground state obtained by exact diagonalization can never correspond to a localized soliton. It will instead contain a superposition of solitons centered around arbitrary $\theta$. This problem can be overcome by superposing a weak symmetry breaking potential to break the degeneracy between states with a different soliton position:
\begin{eqnarray}
\hat{H}_{sb} &=& \hat{H} + \hat{V}_\epsilon \\
\hat{V}_\epsilon &=& \frac{\epsilon}{N^2} \int d\theta \opsid \cos{\theta} \opsi \,.
\end{eqnarray}
The higher $\epsilon$, the deeper the symmetry breaking potential, and the more localized the soliton will be. 

\section{Quantum Phase Transition in the 1D-Bose gas}
The mean field treatment of the attractive 1D Bose gas above has signalled a
quantum phase transition. The degeneration of the Bogoliubov modes at
$\alpha N = 1$ supports the existence of a critical point. Although, by
definition, a phase transition can only occur for infinite $N$,
indications for it should already be visible for large but finite $N$. Here we
will focus on two observations:
\begin{itemize}
   \item[(i)] The one-particle entanglement entropy displays a sharp increase close to the critical point.
   \item[(ii)] The ground state fidelity peaks at the critical point; the height of the peak grows with $N$.  
\end{itemize} 

\subsection{One-Particle Entanglement}
The one particle entanglement entropy is defined as the von Neumann
entropy $S_1 = \tr[\hat\rho_{1\text{p}} \log \hat\rho_{1\text{p}}]$ of the one
particle density matrix $\hat\rho_{1\text{p}}$ of the ground state, obtained by 
singling out one particle and tracing over all $N-1$ other. As long as the
ground  state of the system is well described by a Hartree, i.e. product state, 
$\hat\rho_{1\text{p}}$ describes a pure state; the entanglement entropy vanishes. When the
critical point is approached, collective effects become important. No longer is
the  ground state described by a product state; consequently the entanglement
entropy  increases - a single particle becomes strongly entangled with the rest
of the system.

\begin{figure}[t]
\begin{center}
\includegraphics[width=.75\linewidth,angle=0.]{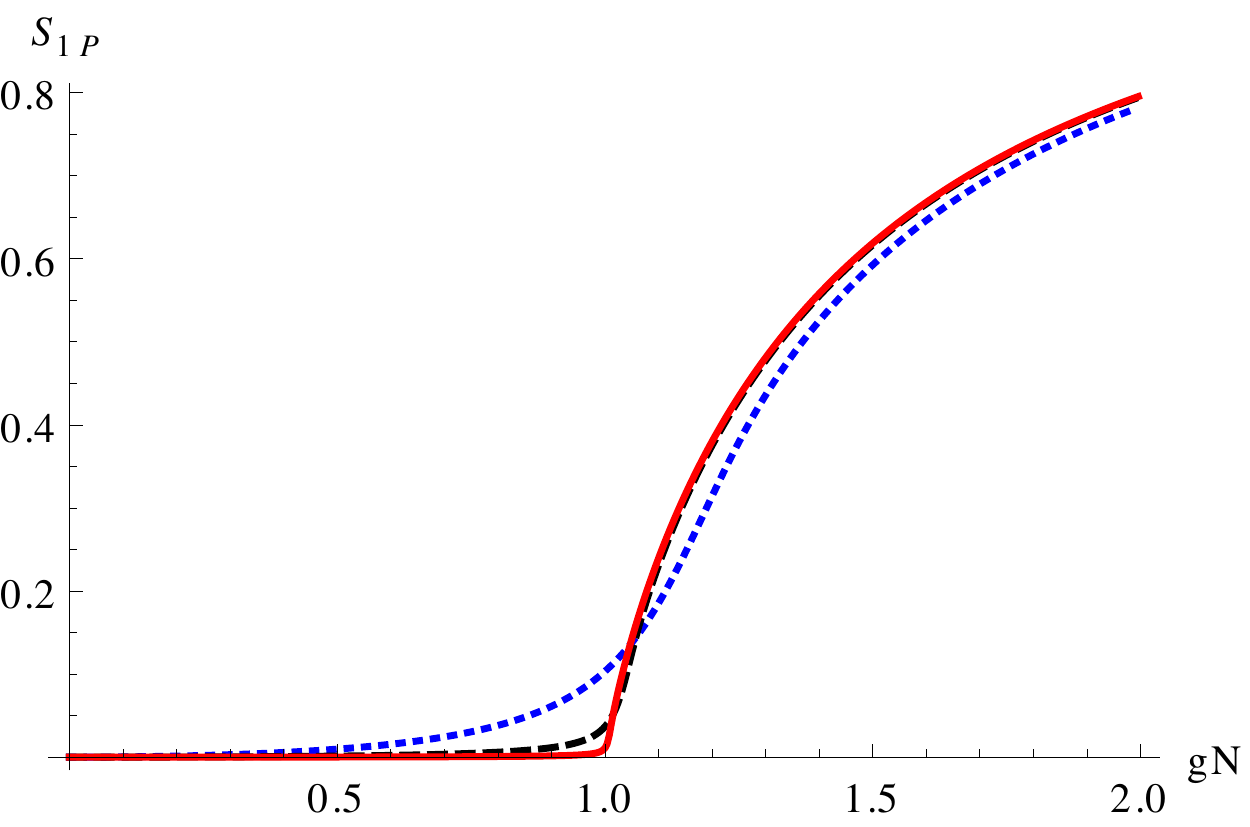}
\end{center}
\caption{One-particle entanglement entropy for $N = 50$ (blue, dotted), $500$ (black, dashed), $5000$ (red, solid). 
}
\label{fig:1pentent}
\vspace{0.5cm}
\end{figure} 

The one particle density matrix is defined via
\begin{equation}
\label{1pdm}
\hat\rho_{1\text{p}} = \tr_{(N-1)\text{P}} \hat\rho = \tr_{(N-1)\text{p}}
\ket{0_{GP}}\bra{0_{GP}} \,,
\end{equation}
or, explicitly, in the one particle momentum eigenbasis
\begin{equation}
\left(\hat\rho_{1\text{P}}\right)_{ij} = \delta_{ij} \sum_{\{n_k\}}
|\alpha_{\{n_k\}}|^2 \frac{n_i}{N} \,.
\end{equation}
Here, $n_k$ is the occupation number of the $k$-th momentum mode and we have used 
\begin{equation}
\ket{0_{GP}} = \sum_{\{n_k\}}\alpha_{\{n_k\}} \ket{\{n_k\}}\,.
\end{equation}
We have plotted the numerically evaluated one particle entanglement as a
function of $\alpha N$ for different $N$ in Fig.\ref{fig:1pentent}. The increase
close to the critical point gets profoundly sharper for larger $N$. Independent 
of $N$, the entropy is bounded by $S_{\rm max} = \log 3$, due to the truncation 
of the one-particle Hilbert space to a three level system.

The entanglement entropy becomes maximal for large $\alpha N$. This, as argued before, is due to the fact that the numerical groundstate is given by a superposition of solitons localized at arbitrary positions.\footnote{This quantum behavior is not expected to survive in the large $N$ limit if the symmetry breaking potential is turned on. In this case, a vacuum is selected which does not mix with translated states (see Appendix). The entanglement entropy is therefore much smaller.}

\subsection{Ground State Fidelity}
Ground State Fidelity (GSF) was introduced in \cite{fidelity} as a characteristic
of a QPT. It is defined as the modulus of the overlap of the exact ground states
for infinitesimally different effective couplings.
\begin{equation}
F(\alpha N, \alpha N+\delta) = |\bra{0_{\alpha N}}0_{\alpha N+\delta}\rangle|
\end{equation}
Far away from the critical point, this overlap will be very close to unity.
For small $\alpha N$, the ground state is dominated by the homogeneous state, and
while coefficients may change slightly, no important effect will be seen. The 
analogous statement holds deep in the solitonic regime. While the shape of the
soliton changes, it will so smoothly; in the infinitesimal limit, the overlap is
one. Right at the critical point, however, the ground state changes in a
\emph{non-analytic} way. The homogeneous state ceases to be the
ground state and becomes an excited state, while the soliton becomes the new
ground state. As energy eigenstates with different eigenvalue are orthogonal,
the ground state fidelity across the phase transition is exactly zero.
\begin{figure}[t]
\centering
\includegraphics[width=.75\linewidth,angle=0.]{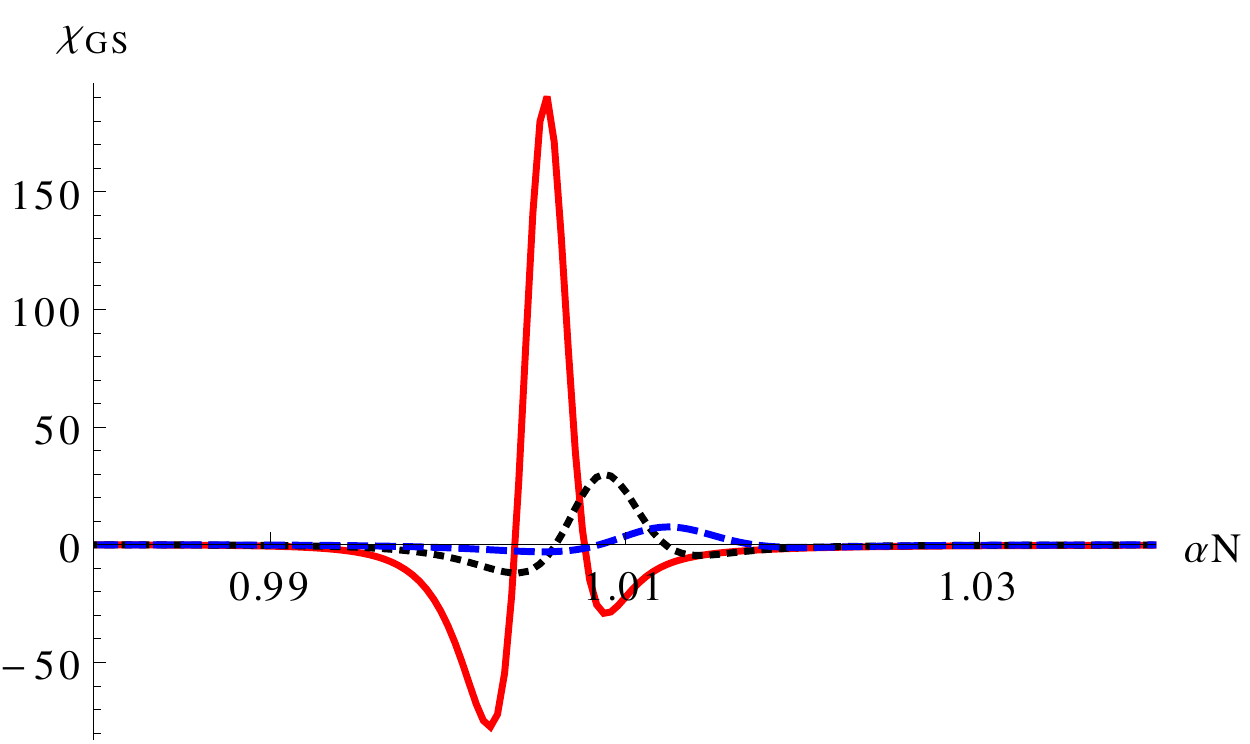}
\caption{Numerical ground state fidelity susceptibility for $N = 3000$ (blue, dashed), $N = 5000$ (black, dotted) and $N = 10000$ (red, solid).
\label{fig:fidelity_num}}
\vspace{0.5cm}
\end{figure}

The GSF has the disadvantage of depending on the arbitrary choice of the small
parameter $\delta$. This can be cured by introducing the fidelity susceptibility
$\chi_{\rm gs}(\alpha N)$ as the second derivative of the GSF.
\begin{equation}
\chi_{\rm gs}(\alpha N) = \lim_{\delta \to 0}\frac{F(\alpha N,\alpha N+\delta)-F(\alpha N,\alpha N-\delta)}{\delta^2} \,.
\end{equation}
It has been shown \cite{yang} that singular behavior of the fidelity
susceptibility directly signals a discontinuity of the first or second
derivative of the ground state energy - a quantum phase transition.

The aforementioned behavior is of course idealized for an infinite system, where
ground state degeneracy and thus level crossing become an exact property. In the
finite $N$ systems we examined numerically, the overlap cannot go to zero,
because there is anticrossing which allows the energy levels to degenerate
only for $N \rightarrow \infty$.

Still we can observe a drop in the fidelity which 
deepens with $N$ but is of magnitude much smaller than $1$ for all $N$ we
were able to simulate. The fidelidy susceptibility as obtained from the exact
diagonalization is plotted in Fig.\ref{fig:fidelity_num} for different $N$.
In the limit $N \to \infty$, we expect a behavior $\chi_{\rm gs} \to -\delta''(\alpha N-1)$. This tendency can be clearly observed.
Both the negative and the positive peak move towards $\alpha N = 1$, they become narrower, and their modulus diverges with growing $N$.

\section{Fluctuation Entanglement}
We will now consider the entanglement between the fluctuation 
$\delta\hat a_k = \hat a_k - a_k^\text{c}$ of a given original momentum mode and
the fluctuations of the rest of the system. The motivation for studying this
quantity is twofold. We imagine, that an external observer would couple linearly
to the bosonic field (so that the situation has some minimal resamblance with
the gravity case). It has been pointed out \cite{Anglin:1995pg} that
for such a coupling, field  values (or their Fourier components) will be the
environment-selected pointer states\footnote{Pointer states denote those states
that are stable with respect to interactions with the environment and therefore
correspond to classically observable states.} and not localized single particle
states. This leads us to consider the entanglement of a momentum mode, rather than single-particle entanglement, as a measure of relevant quantum correlations of the given state. Furthermore, the observer couples to the original field $\hat a_k$ and hence its
fluctuations as opposed to coupling to the Bogoliubov modes $\hat b_k$.

More technically speaking, the quantity we calculate is the von
Neumann entropy of the reduced density matrix for a given $\delta\hat a_k$
\begin{equation}
	(\delta\rho_k)_{n m} = \tr_{\text{modes }k' \neq k}\left[ \rho \,
	(\delta\hat a_k^\dagger)^m|0^\text{c}\rangle\langle 0^\text{c}| (\delta\hat
	a_k)^n \right]
\end{equation}
where $|0^\text{c}\rangle$ denotes the state that would be observed classically.

Fluctuation entanglement provides a measure for the quantum correlations between a single momentum mode with the rest of the system. It hence gives a direct handle of the quantumness of our ground state as measured by an outside observer if coupled linearly to the field. Note also that due to the fact that we are considering a closed system, the fluctuation entanglement is exactly equivalent to the Quantum Discord introduced in works \cite{discord_entanglement} as a measure of quantumness.

\subsection{Calculation in the Bogoliubov Approximation}
In order to calculate the fluctuation entanglement in the Bogoliubov case, note
that the sought-for density matrix is Gaussian\footnote{A density matrix is
called Gaussian, when its Wigner function $W(\alpha, \alpha^\star) =
\tfrac{1}{\pi^2} \int d^2 \beta \exp(-i \beta \alpha^\star - i \beta^\star
\alpha) \tr[\rho \exp(i \beta a^\dagger + i \beta^\star a)]$ is Gaussian}. The
ground state in terms of $\hat b_k$ is Gaussian and the Bogoliubov transformation amounts to squeezing - which leaves a Gaussian state Gaussian. Also integrating out modes in a Gaussian state does not change this property. Hence the reduced density matrix in terms of $\delta\hat a_k$ must have the form
\begin{equation}
	\rho_k=C_k\exp\left\{-\lambda_k\left( \delta\hat a^\dagger_k
	\delta\hat a_k+-\frac{1}{2}\tau_k\left[\delta\hat a^\dagger_k \delta\hat
	a^\dagger_k + \delta\hat a_k \delta\hat a_k\right]\right)\right\},
\label{density_matrix}	
\end{equation}
with real coefficients $\lambda_k$ and $\tau_k$ and normalization $C_k$ such
that $\tr \rho_k = 1$. This density matrix has a von Neumann entropy 
\begin{equation}
	S_k=\frac{\lambda_k\sqrt{1-\tau_k^2}}{2}\left( \coth{\frac{\lambda_k\sqrt{1-\tau_k^2}}{2}}-1 \right)-\ln\left(1-e^{-\lambda_k\sqrt{1-\tau_k^2}}\right)
\end{equation}

We can fix the unknown coefficients by imposing
\begin{align}
	\label{dons-cons}
	\langle\psi| \delta\hat a^\dagger_k \delta\hat a_k|\psi\rangle &=
	\tr[\rho_k \delta\hat a^\dagger_k \delta\hat a_k] &
	&\text{and}&
	\langle\psi|\delta\hat a_k \delta\hat a_k|\psi\rangle &=
	\tr[\rho_k \delta\hat a_k \delta\hat a_k],
\end{align}
where $|\psi\rangle$ is the groundstate of the Bogoliubov modes.

\subsection{Homogenous Phase}
In the homogenous case, imposing (\ref{dons-cons}) and evaluating the left hand
side by inserting the Bogoliubov transformation (\ref{bog-hom}) leads
to
\begin{align}
	\lambda_k&=\ln\left(\frac{u_k}{v_k}\right)^2,&
	\tau_k&=0&&\text{and} & 
	C_k&=1/u_k^2.
\end{align}
Thus, the fluctuation entanglement entropy is
\begin{equation}
	S_k=u_k^2 \ln u_k^2 -v_k^2 \ln v_k^2.
\end{equation}
The entanglement of the first momentum mode $S_1$ diverges near the critical
point $\alpha N=1-\delta$ as
\begin{equation}
	S_1\approx 1-\ln (4)-\frac{1}{2}\ln \delta.
\end{equation}
A similar divergence of an entanglement entropy has been pointed out in spin
chain (and analogous) systems undergoing a phase transition
\cite{vidal_osborne}. In contrast to these cases however, where the
entanglement is between nearest neighbour sites, the diverging
entanglement in our case is between different low-momentum modes and not between
localized sites. So one may say, that the entanglement in our case is long-range.
Furthermore it should be noted, that the entanglement of the higher modes $|k|>1$ stays finite near the critical point, 
showing that the diverging entanglement is an infrared effect, which can be expected to be independent of short distance physics.

\subsection{Solitonic Phase}
\begin{figure}[t]
  \centering
  \includegraphics[width=.75\linewidth,angle=0.]{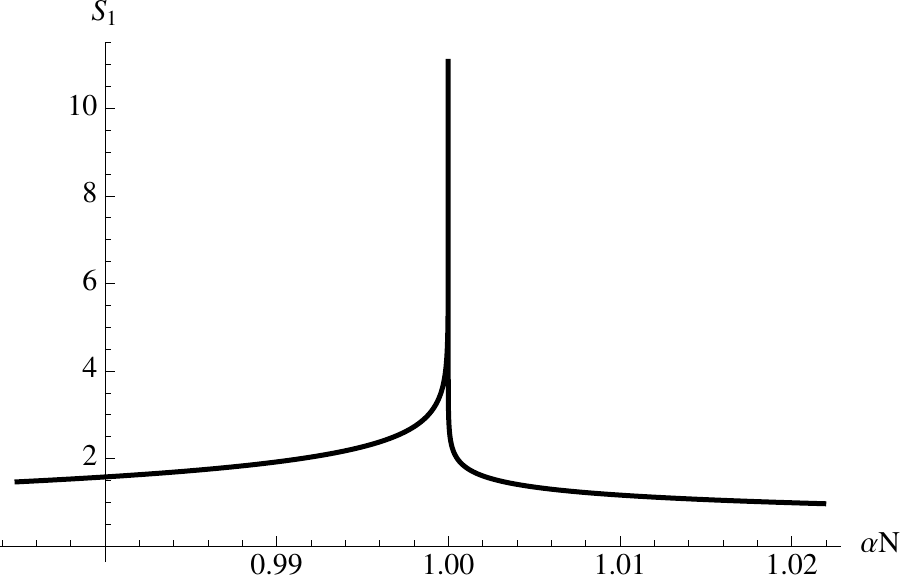}
  \caption{Analytical fluctuation entanglement}
    \label{bogoliubov3}
\end{figure}

The relevant expectation values in the Bogoliubov ground state are given by
\begin{align}
	\langle\psi|\delta\hat a^\dagger_m \delta\hat a_n|\psi\rangle &= 
	\sum_k \left(\int e^{im\theta} v_k(\theta)d\theta\right)\left(\int
	e^{-in\theta} v_k(\theta)^\star d\theta\right),\\ 
	\langle\psi|\delta\hat a_m \delta\hat a_n|\psi\rangle &= 
	\sum_k \left(\int e^{-im\theta} u_k(\theta)d\theta\right) \left(\int
	e^{-in\theta} v_k(\theta)^\star d\theta\right).
\end{align}
It can be checked that close to the phase transition the first excited mode 
gives the leading contribution to the aforementioned entanglement entropy. 
The quantities $\langle\psi|\delta\hat a^\dagger_1 \delta\hat a_1|\psi\rangle$
and $\langle\psi|\delta\hat a_1 \delta\hat a_1|\psi\rangle$ can be obtained  by
numerical integration. The parameters $\lambda,\tau$  of the reduced gaussian
density matrix can then be determined. The final von Neumann entropy
again shows a divergence\footnote{The shape of the divergence obtained
by numeric integration seems to be consistent with a logarithm with a
coefficient close to $0.33$ near the phase transition.} close to the phase
transition. Fig.\ref{bogoliubov3} shows the fluctuation entanglement obtained in the Bogoliubov
approximation on both sides of the phase transition.
 
\begin{figure}[t]
  \centering
  \includegraphics[width=.75\linewidth,angle=0.]{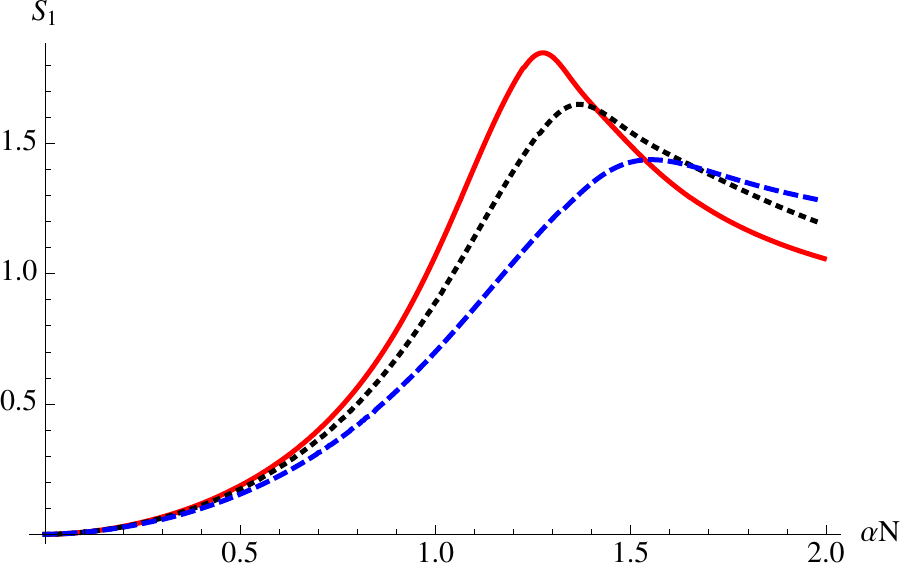}
  \caption{Numerical fluctuation entanglement for $N = 15$ (blue, dashed), $N = 20$ (black, dotted) and $N = 25$ (red, solid).}
    \label{bogentnum}
\end{figure}

\subsection{Numerical Treatment}
As discussed before, for any given finite size system, the Bogoliubov
approximation should not be trusted close to the critical point. Therefore it is
important to study the exact behavior of finite size systems numerically, in
order to substantiate the claim that the fluctuation entanglement entropy
becomes large.

Within an exact treatment this quantity is considerably more difficult to
extract, because in contrast to the Bogoliubiv analysis, one does not have
direct access to a ``classical background'' which one could use to disentangle
classical correlations. Instead, the seperation can be obtained through the
following procedure.

Since all numerical solutions are obtained for a fixed particle number $N$, the field expectation value in the exact ground state $\Ngs$ will necessarily vanish, $\bra{0_N} \opsi \ket{0_N} = 0$. Obviously, the ground state $\Ngs$ can hence never correspond to the classical (coherent) state with a wave function corresponding to the soliton solution of the Gross-Pitaevskii equation, $\bra{\psi_\text{cl}}\opsi\ket{\psi_\text{cl}} = \Psi_\text{GP}(\theta)$. In order to define a mapping from $\Ngs$ to $\ket{\psi_\text{cl}}$, we numerically search for the coherent state $\ket{\alpha}$ with maximal overlap with $\Ngs$. This state is expected to be annihilated by the perturbations of the Gross-Pitaevskii ground state,
\begin{equation}
	\label{gpgs} \delta\hat{a}_k \ket{\alpha_k} = 0 \,,
\end{equation}
where
\begin{equation}
	\label{darel}
	\delta\hat{a}_k = \hat{a}_k - c_k
\end{equation}
and the $c_k$ are the Fourier coefficient of $\Psi_\text{GP}(\theta)$. From Eq.(\ref{gpgs}) it directly follows that $\alpha_k = c_k$.
There is now an obvious measure of correlations which excludes those of the Gross-Pitaevskii background: The entanglement entropy of the $\delta\hat{a}_1$ modes, described by the density matrix
\begin{equation}
	\label{dadens}
	(\tilde\rho_1)_{kl} = \tr\left[\rho \ket{\delta l_1}\bra{\delta k_1}\right] \,.
\end{equation}
Now, $\ket{\delta l_1}$ denotes the eigenstate of $\delta\hat{N}_1 = \delta\hat{a}^\dagger_1 \delta\hat{a}_1$ with eigenvalue $\delta l_1$. Eq. (\ref{dadens}) directly corresponds to the density matrix (\ref{density_matrix}) in the Bogoliubov approximation. Using the relations (\ref{gpgs}) and (\ref{darel}), it can be directly recast to take on the form
\begin{equation}
(\tilde\rho_1)_{kl} = \tr\left[\rho \frac{(\hat{a}_1^\dagger -
\alpha_1^*)^l}{\sqrt{l!}}\ket{\alpha_1}\bra{\alpha_1}\frac{(\hat{a}_1 - \alpha_1)^k}{\sqrt{k!}}\right] \,,
\end{equation}
which, by the definition of a coherent state, can be straightforwardly evaluated.

The resulting fluctuation entanglement is shown in Fig.\ref{bogentnum} for
different particle number. It has a clear maximum at the
would--be--phase--transition. The maximum value becomes larger and the peak
narrower with increasing particle number, so the divergence in the
Bogoliubov case seems a plausible limit. 
The fact that at $\alpha N = 2$, the fluctuation entanglement is still quite high is not surprising. Only in the limit $N \to \infty$ do we expect to see the behavior of the Boguliubov analysis. This is supported by the fact that a decrease is observed for increasing $N$, as well as for stronger localization potentials.

\section{Conclusions and Outlook}

In this paper, we have considered properties of the  1+1-dimensional
attractive Bose gas around its critical point. By analyzing important
indicators for QPTs, we provided further evidence that a tuning of the effective coupling $gN$
leads  to a phase transition in the system. More importantly, we have shown
that quantum correlations become very important close to the critical
point - contrary  to the naive intuition that at sufficiently large particle
number, systems should behave  approximately classical. We have also pointed
out that the quantum entanglement of the bosons close to the critical point is
``long range'' - in contrast to the observations in spin-chain systems that
display nearest neighbour entanglement at criticality. 

The motivation for our study of this model system, however, was the conjecture
that black holes are bound states of a large number of weakly interacting
gravitons. It has been claimed that the graviton condensates behave
significantly different with respect to the semiclassical black hole analysis
due to their being at a quantum critical point. It was argued that criticality
allows quantum effects to only be suppressed by the perturbative coupling
$\alpha_g \sim 1/N$. If the qualitative insights from our simple toy model are
valid for graviton condensates our results can back up several of the
claims. We can argue that quantum effects become important for attractive Bose
condensates at their critical point - even though the perturbative coupling is
very small. And that the entanglement of the true state is long range -
consistent with the notion of a condensate of gravitons of
wavelength comparable to the Schwarzschild radius. This would imply that for a
black hole, the semiclassical treatment with a background geometry that obeys
classical general relativity and quantization of fields on top of this rigid
background becomes invalid much earlier than what the standard lore tells.
Although curvature invariants in the horizon region of large Schwarzschild black
hole are small, the semiclassical treatment is not applicable.
Instead, quantum correlations in the graviton bound state become relevant.
Importantly, our results point in the direction that the physics is dominated by
large wavelengths. Therefore the description of black holes as graviton
condensates has the attractive feature of being independent of the ultraviolet
completion of gravity. The only requirement being that the low
energy theory resembles perturbatively quantized Einstein theory with a massless
spin two graviton.

The 1+1-dimensional Bose gas can indeed capture quite a few of the
intriguing features  of black holes and their possibly quantum nature.
To understand in more detail time dependent features, such as Hawking evaporation, resolutions 
of the information paradox or scrambling, the implementation of dynamical methods will be amongst 
the aims of immediate future work. This will then necessarily also address possible couplings 
to external systems in order to be able to model the evaporation process. Working with more 
spatial dimensions
may prove feasible to model the collapse induced by Hawking evaporation. This could alternatively 
be achieved by considering couplings that show further resemblance with gravitational self-
interactions. Steps in these directions also include generalizations to non-number conserving, and ultimately relativistic theories.

Instabilities can in turn be countered by adding repulsive interactions that dominate
at  very short scales. Stable configurations of that sort would correspond to extremal
black  holes. Their properties also provide a vast playground for future 
investigation.

\subsection{Acknowledgements}
We are grateful to Gia Dvali and Cesar Gomez for initiating our work and many useful discussions.
Furthermore, we thank Felix Berkhahn, Sarah Folkerts, Andr\'e Fran\c{c}a, Cristiano Germani, Florian Niedermann and Tehseen Rug for discussions and comments. Finally, we would like to thank Francesco Piazza and Richard Schmidt for sharing their insights on the physics of Bose-Einstein condensates and ultracold atoms.
The work of all authors was supported by the Alexander von Humboldt foundation.

\appendix
\section{Appendix}

\subsection{Spontaneous symmetry breaking in finite volume}
Standard lore states that there can be no spontaneous symmetry breaking in finite volume with a finite number of fields. This may seem puzzling since we claim here explicitly that our ground state is a localized soliton for $gN>1$, clearly signaling spontaneous breaking of translation invariance.
This puzzle is resolved by noticing that the mean field approximation can only be exact in the limit $N \to \infty$.

To illustrate this, consider a number conserving Hamiltonian containing a finite number $2m$ of fields in the potential.
Mean field approximation becomes exact if the corresponding state can be approximated arbitrarily well by a Hartree-type state, i.e.
\begin{equation}
	\ket{\psi_0}=\otimes \sum_{k}c_k \ket{k}\,.
\end{equation}
A translated state is given by
\begin{equation}
	\ket{\psi_{\delta \theta}}=\otimes \sum_{k}c_k e^{i k \delta \theta} \ket{k}\,.
\end{equation}
Now it can be seen that spontaneous symmetry breaking is indeed possible as we take the limit $N\rightarrow \infty$. It can be shown that
\begin{align}
	\lim_{N\rightarrow \infty}\left|\braket{\psi_0}{\psi_{\delta
	\theta}}\right|^2 &=\lim_{N\rightarrow \infty}\left| \sum_{k}c_k e^{i k \delta
	\theta} \right|^{2N}\,,\\ 
	\lim_{N\rightarrow \infty}\left|\bra{\psi_0}H \ket{\psi_{\delta
	\theta}}\right|^2 &\leq \lim_{N\rightarrow \infty} C N^{2m}\left| \sum_{k}c_k
	e^{i k \delta \theta} \right|^{2(N-m)}\,.
\end{align}
In both expressions the right hand side vanishes as long as $c_k \neq 0$ for at least two different k. In other words, in the large $N$ limit localized objects, localized at different points are orthogonal and, furthermore, do not mix under time evolution.

In the proof for absence of finite volume symmetry breaking enters the assumption that expectation values of composite operators made out of the fields are finite. This assumption breaks down in the large $N$ limit. While this implies unboundedness of the energy in this limit, it is nothing to worry about. The energy diverges linearly in $N$ and hence the energy per particle remains bounded, reminiscent of a thermodynamic limit. 
 
\end{document}